\documentclass[prl,a4paper,superscriptaddress,twocolumn,showpacs,amsmath,amssymb,floatfix,amstex]{revtex4}

\usepackage{psfrag}

\usepackage[dvips]{graphicx}
\usepackage{bm,pstricks,longtable}

\input{epsf}

\begin{document}

\title{Delocalization transition for the Google matrix}
\author{Olivier Giraud}
%\homepage[]{http://www.quantware.ups-tlse.fr}
\affiliation{\mbox{Universit\'e de Toulouse, UPS,
Laboratoire de Physique Th\'eorique (IRSAMC), F-31062 Toulouse, France}}
\affiliation{\mbox{CNRS, LPT (IRSAMC), F-31062 Toulouse, France}}
\author{Bertrand Georgeot}
%\homepage[]{http://www.quantware.ups-tlse.fr}
\affiliation{\mbox{Universit\'e de Toulouse, UPS,
Laboratoire de Physique Th\'eorique (IRSAMC), F-31062 Toulouse, France}}
\affiliation{\mbox{CNRS, LPT (IRSAMC), F-31062 Toulouse, France}}
\author{Dima L. Shepelyansky}
%\homepage[]{http://www.quantware.ups-tlse.fr}
\affiliation{\mbox{Universit\'e de Toulouse, UPS,
Laboratoire de Physique Th\'eorique (IRSAMC), F-31062 Toulouse, France}}
\affiliation{\mbox{CNRS, LPT (IRSAMC), F-31062 Toulouse, France}}

%\date{\today}
\date{March 30, 2009}

\begin{abstract}
We study the localization properties of eigenvectors of the Google matrix, 
generated both from the World Wide Web and from the Albert-Barab\'asi model
of networks.  We establish the emergence of a delocalization phase
for the PageRank vector when network parameters are changed. In the phase
of localized PageRank, a delocalization takes place in the complex plane of 
eigenvalues of the matrix, leading to delocalized relaxation modes.  We argue that the
efficiency of information retrieval by Google-type search is strongly affected in
the phase of delocalized PageRank.
\end{abstract}

%89.20.Hh  	World Wide Web, Internet
%89.75.Hc       Networks and genealogical trees 
%05.40.Fb       Random walks and Levy flights
%72.15.Rn Localization effects (Anderson or weak localization) 
%87.23.Ge       Dynamics of social systems
% 64.60.Fr      Equilibrium properties near critical points, critical exponents

\maketitle

The World Wide Web (WWW) is an enormously large network
 with about $10^{11}$ webpages all over the world.
Information retrieval in such a huge database is
therefore a formidable task.  An efficient method to
search this database, known as the PageRank Algorithm (PRA), 
was put forward by Brin and Page
\cite{brin} and formed the basis of the Google
search engine, by far the most popular one.
The PRA is based on the construction of the Google matrix ${\bf G}$ which 
sums up the network structure 
in a tractable way
and can be written as (see e.g. 
\cite{googlebook} for details)
\begin{equation}
{\bf G}=\alpha {\bf S}+(1-\alpha) {\bf E}/N.
\label{eq1}
\end{equation}
The matrix ${\bf S}$ is constructed from the adjacency matrix of the network.  
For a directed network of $N$ nodes,
the $N\times N$
adjacency matrix ${\bf A}$ is defined by $A_{ij}=1$ if there is a link from
node $j$ to node $i$, and $A_{ij}=0$ otherwise. For networks with undirected 
links, ${\bf A}$ is a real symmetric matrix.  However, the WWW corresponds to a network with
directed links and here ${\bf A}$ is not
symmetric. Matrix $S_{ij}$ is built from ${\bf A}$ by normalizing each nonzero
column through $S_{ij}=A_{ij}/\sum_k A_{kj}$ and
replacing by $1/N$ the elements of columns with
only zero elements.  The matrix ${\bf S}$ can be viewed as the mathematical description
of a surfer on the network. 
At each iteration he leaves a node by 
randomly choosing an outgoing link
with equal probability, and in the absence of such links
he goes to an arbitrary node at random.  The Google matrix ${\bf G}$ defined 
by Eq.(\ref{eq1}) (with matrix ${\bf E}$ such that all $E_{ij}=1$)
can be interpreted as a modification of ${\bf S}$ where
 with finite probability $1-\alpha$ the surfer might jump to another
node at random. Usually the PRA uses
$\alpha=0.85$ and we concentrate our studies on this case.

The matrix ${\bf G}$ has only one maximal eigenvalue $\lambda=1$.  
The corresponding PageRank
eigenvector with components $p_j$ 
gives the stationary distribution of the random surfer over the network.  
All $p_j$ are 
positive real numbers normalized by $\sum p_j=1$. All nodes in the WWW 
can be ordered by decreasing $p_j$ values
and thus this PageRank vector is of primary importance for ordering of 
websites and information retrieval.
The vector can be found by iterative applications
of ${\bf G}$ on an initial random vector.  This PRA works efficiently
due to the relatively small average number of links in the WWW.   The WWW
is indeed described by a very sparse  adjacency matrix ${\bf A}$, 
with only about ten nonzero entries per column.

Numerical studies of the PageRank vector for large subsets of the WWW
have shown that it is satisfactorily described by an algebraic 
decay $p_j\sim 1/j^{\beta}$ where $j$ is the ordered index, and thus
the number of nodes $N_n$ with PageRank $p$
scales
as $N_n\sim 1/p^{\nu}$ with numerical values $\nu = 1+1/\beta \approx 2.1$ 
and $\beta\approx 0.9$ \cite{donato}.  This implies that
the PageRank vector is not ergodic, displaying certain localization properties
over specific sites of the network.  The localization properties of eigenvectors
of real symmetric matrices describing various complex networks have been studied recently.
For systems of small-world type it was shown that eigenvectors display
a transition from localized to delocalized states when the density
of long-range links is changed \cite{giraud,berkovits}.  Such delocalization transition
has certain similarities with the Anderson transition for waves in systems with disorder
\cite{anderson}.
More specific studies were
performed for the symmetric adjacency matrix of the Internet network, showing
that the localization of eigenvectors strongly depends on the eigenvalue location 
in the spectrum, and allows to identify isolated communities \cite{maslov}.  
The global localization properties averaged over the spectrum were also recently considered in 
\cite{baowen} for various undirected networks.
The studies above were performed for symmetric adjacency matrices of undirected networks,
characterized by real eigenvalues.  In contrast, the Google matrix is constructed on
the basis of directed links, and thus its spectrum is generally complex.  
We note that the case of complex spectra in quantum mechanics
was studied in relation to 
poles of scattering problems (see e.g. \cite{fyodorov}) 
but it remains less explored than the case of real spectra.  

In this Letter, we study the localization properties of the Google matrix ${\bf G}$
 for models of realistic directed networks and actual subsets of the WWW.  
We characterize the properties of right eigenstates 
$\psi_i$ (${\bf G} \psi_i=\lambda_i \psi_i$) as a function of the complex eigenvalue 
$\lambda$.  Special emphasis is given to the properties of the PageRank vector, which
is of great importance for the Google search.  Our findings show that eigenstates
with complex $\lambda$ are generally delocalized over the whole network.  At the same time,
the PageRank vector may be localized or delocalized depending on the properties 
of the network.  Such delocalization may seriously affect the efficiency of the 
ranking through the PRA.  We note that the PRA has recently found new types of
applications e.g. for academic ranking from citation networks \cite{redner}.  It
is rather probable that the PRA will find broad application for classification in various
types of complex networks \cite{dorogovtsev} and hence, the understanding of
global properties of the Google matrix becomes very important.

\begin{figure}[htbp]
\begin{center} 
%\vspace{0.5cm}
%%\includegraphics[width=.95\linewidth]{fig1a.eps}\\
%%\vspace{-0.6cm}
%%\includegraphics[width=.95\linewidth]{fig1b.eps}\\
%%\vspace{-0.5cm}
%%\includegraphics[width=.95\linewidth]{fig1c.eps}
\includegraphics[width=.95\linewidth]{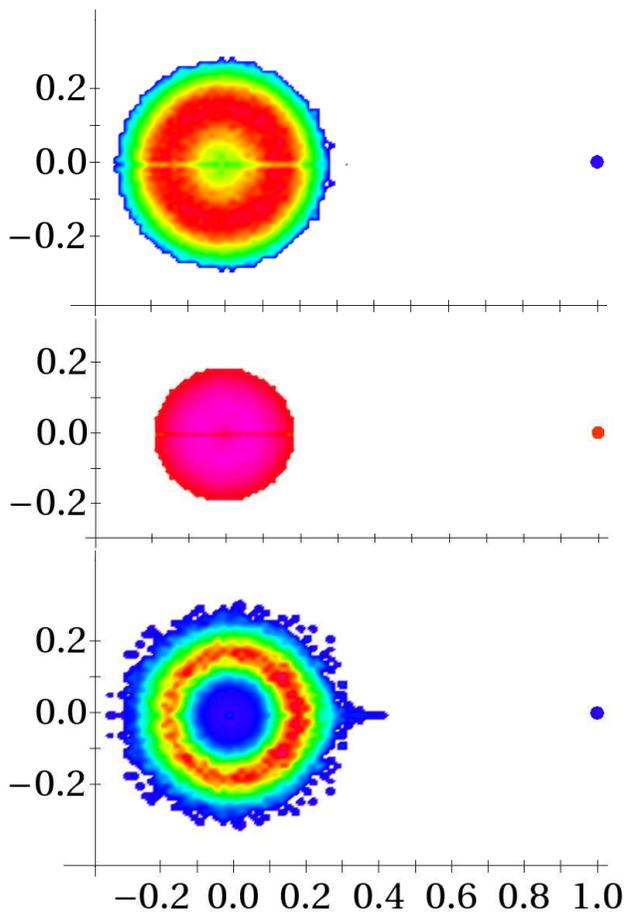}
\end{center} 
 \vglue -0.3cm
\caption{(Color online) Distribution of eigenvalues $\lambda_i$ of Google matrices
in the complex plane. Color is proportional to the IPR $\xi$ of the
associated eigenvector $\psi_i$. Top panel: AB model with $q=0.1$ 
for $N=2^{14}$, $N_r=5$ random realizations, $\xi$ varies from
$\xi=32$ (blue/black) to $\xi=1656$ (red/grey); middle panel: same with $q=0.7$,
 $\xi$ varies from
$\xi=1169$ (red/grey) to $\xi=3584$ (purple/dark grey); bottom panel:
data for a University network (Liverpool J. Moores Univ. - LJMU) 
with $N=13578$ and $N_r=5$ (see text),
 $\xi$ varies from
$\xi=7$ (blue/black) to $\xi=1177$ (red/grey).
}
\label{fig1}
\end{figure}

To generate Google matrices ${\bf G}$ we use data from real subsets of the WWW,
namely University networks taken from \cite{uni}.  In addition, we 
generate networks with directed links using the Albert-Barabasi (AB) procedure
\cite{albert} to construct the associated ${\bf G}$ matrix. AB networks are built
by an iterative process. Starting from $m$ nodes, at each step $m$ links are added to the
existing network with probability $p$, or $m$ links are rewired with probability $q$,
or a new node with $m$ links is added with probability $1-p-q$.  In each case the end
node of new links is chosen with preferential attachment, i.e. with probability 
$(k_i+1)/\sum_j(k_j+1)$ where $k_i$ is the total number of incoming and outgoing links
of node $i$.  This mechanism generates directed networks having the small-world and scale-free
properties, depending on the values of $p$ and $q$.  The results we display are averaged
over $N_r$ random realizations of the network to improve the statistics.
 In our studies we chose $m=5$, $p=0.2$ and two values of $q$ corresponding to scale-free
 ($q=0.1$) and exponential ($q=0.7$) regimes of link distributions (see Fig.~1 in \cite{albert}
for undirected networks).  For our directed networks at $q=0.1$, we find 
properties close to the behavior for the WWW
with the cumulative distribution of ingoing
links showing algebraic decay $P_c^{in}(k) \sim 1/k$
and average connectivity $\langle k \rangle \approx 6.4$. 
For $q=0.7$ we find $P_c^{in}(k) \sim \exp(-0.03k)$ and $\langle k \rangle \approx 15$.
For outgoing links, the numerical data are compatible with an exponential decay in both cases
with $P_c^{out}(k) \sim \exp(-0.6 k)$ for $q=0.1$ and $P_c^{out}(k) \sim \exp(-0.1 k)$
for $q=0.7$.  We checked that small variations of parameters $m, p,q$ near the chosen values
do not qualitatively affect the properties of ${\bf G}$ matrix.  

To characterize localization properties of eigenvectors $\psi_i$, 
we use the Inverse Participation Ratio
(IPR) defined by $\xi=(\sum_j |\psi_i(j)|^2)^2/\sum_j|\psi_i(j)|^4$. It gives the effective
number of nodes on which an eigenstate is localized.  In Fig.~\ref{fig1} we show
the distribution of eigenvalues together with the IPR
for the AB model and the WWW.  In the latter case, to improve the statistics we randomize
the links, keeping fixed the number of links at any given node as proposed in \cite{maslov2}. 
In all cases the spectrum consists of an isolated eigenvalue $\lambda=1$ together with an
approximately circular distribution centered at $\lambda=0$ (a significant fraction 
of about 30-50\% states has $\lambda= 0$).  In all three cases there are circular rings of
states with high IPR indicating that in this region the states become delocalized
in the limit of large matrix sizes.  The delocalized domain is largest for AB model
at $q=0.7$, where almost all states have high IPR, including the PageRank vector.
By contrast, at $q=0.1$ the PageRank has small IPR while large IPR appear only
in a ring centered at $\lambda=0$.  We observe a similar behavior for the WWW data
where the ring of delocalized states is narrower and the PageRank has even smaller IPR.

\begin{figure}[htbp]
\begin{center} 
\vspace{0.1cm}
\includegraphics[width=.95\linewidth]{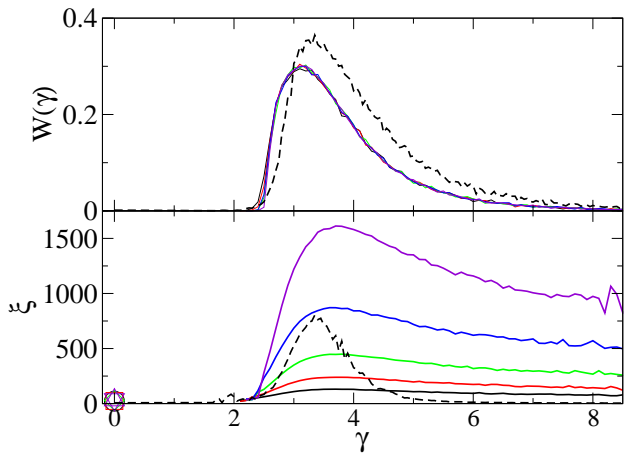}
\end{center} 
\vglue -0.4cm
\caption{(Color online) Normalized density of states $W$ (top panel) and IPR (bottom panel)
as a function of $\gamma$.  Data for AB model with $q=0.1$ are shown by full curves
with from bottom to top $N=2^{10}$($N_r=100$) (black),  
$2^{11}$($N_r=50$) (red), $2^{12}$($N_r=20$) (green), $2^{13}$($N_r=10$) (blue), 
$2^{14}$($N_r=5$) (violet).  Symbols give the PageRank value of $\xi$ in the same order:
circle, square, diamond, triangle down and triangle up.  All curves coincide
on the top panel.  Dashed curves show the data from the WWW
(LJMU network, parameters of Fig.~\ref{fig1}).
}
\label{fig2}
\end{figure}

\begin{figure}[htbp]
\begin{center} 
\vspace{0.6cm}
\includegraphics[width=.95\linewidth]{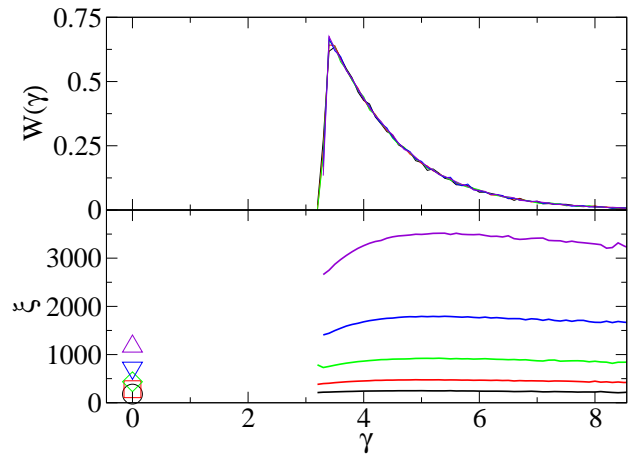}
\end{center}
 \vglue -0.4cm
\caption{(Color online) Same as in Fig.~\ref{fig2} for AB model at $q=0.7$.}
\label{fig3}
\end{figure}

In Figs.~\ref{fig2}-\ref{fig3} we study the dependence on system size $N$. We computed the normalized
density of states $W(\gamma)$ ($\int_0^{\infty} W(\gamma) d\gamma=1$) 
where $\gamma=-2\ln|\lambda|$ is the relaxation rate to the 
equilibrium PageRank state. For AB model in both cases the density
$W(\gamma)$ is independent of system size, showing that we have reached the asymptotic regime
of large networks.  The characteristic features of the density 
are the appearance of a gap between $\gamma=0$ and 
$\gamma=\gamma_c\approx 2-3$, 
followed by a sharp increase with a maximum around 
$\gamma\approx 3-4$ and a slow decrease for larger $\gamma$.  The three models have
a similar structure of $W(\gamma)$, with $\gamma_c$ being not very sensitive to the value
of $\alpha$.  We note that the presence of $\alpha$ in Eq.~(\ref{eq1}) ensures that 
$\gamma_c \geq \gamma_{\alpha}=2| \ln \alpha|$ \cite{googlebook}. For $\alpha=0.85$ this gives 
$\gamma_{\alpha}\approx 0.33$, that is significantly smaller than the numerical value of $\gamma_c$.
This means that all three models have an intrinsic gap that explains the stability of $\gamma_c$
to variations of $\alpha$.  It is known that for WWW networks usually $\gamma_c=\gamma_{\alpha}$.
Indeed, we found that for University networks taken by us from \cite{uni} most often this relation
was approximately satisfied (including for LJMU).  However, randomization of links 
following the procedure of \cite{maslov2} generally increases the size of the gap (see Fig.~\ref{fig1}).  
In order to test the effect of a smaller gap on our results, we also considered a modification of the AB
model where nodes are labeled by an additional ``color'' index, which leads to appearance
of additional eigenvalues in the gap.  This model gives qualitatively similar results
to the models presented here and will be discussed elsewhere. 

While in Figs.~\ref{fig2}-\ref{fig3} $W(\gamma)$ is not sensitive to matrix size,
the IPR clearly grows with $N$ for $\gamma >\gamma_d$, where $\gamma_d$ can
be viewed as a delocalization edge in $\gamma$.  For AB model at $q=0.7$, $\gamma_d =0$
since even the PageRank IPR grows with $N$.  By contrast, for $q=0.1$, 
the PageRank stays constant and $\gamma_d$ is close to
but larger than $\gamma_c\approx 2$.  Data from WWW show a similar behavior of IPR
for fixed matrix size $N$.  

\begin{figure}[htbp]
\begin{center} 
%\vspace{0.5cm}
\includegraphics[width=.95\linewidth]{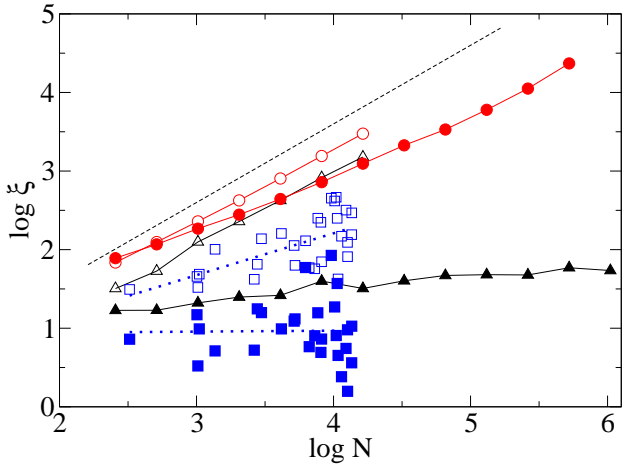}
\end{center} 
\vglue -0.4cm
\caption{(Color online) Dependence of $\xi$ on matrix size $N$ for AB model 
at $q=0.1$ (triangles), $q=0.7$ (circles), and for WWW data without randomization
(squares). Full symbols are for PageRank $\xi$ values, 
empty symbols are for eigenvectors with
$3<\gamma<4$ (AB model) or for the 10 eigenvectors with highest $\xi$ and
$\gamma < 10 $ (WWW data). For AB model $N_r$ is as in Fig.~\ref{fig2} and
$N_r=5$ for $N> 2^{14}$ (statistical error bars are smaller 
than symbol size).
Dotted blue lines give linear fits of WWW data, with
slopes respectively 0.01 and 0.53. Upper dashed line indicates the slope 1.
Logarithms are decimal.}
\label{fig4}
\end{figure}

A detailed analysis of dependence of IPR on $N$ is shown in Fig.~\ref{fig4}, for PageRank
and bulk states with $\gamma > \gamma_c$.  
For  bulk states we find that IPR grows with $N$ as
 $\xi \sim N^{\mu}$ with $\mu \approx 0.9$ (AB model) and $\mu \approx 0.5$  (WWW data).
WWW data in Fig.~\ref{fig4} are taken from actual links 
of various University networks 
without any randomization, which explains a stronger dispersion of data (largest 
not randomized case $N=13578$ corresponds
to the network LJMU used in Figs.~\ref{fig1}-\ref{fig2}).  
The data definitely show that delocalization
takes place in the bulk states.  By contrast, 
the PageRank remains localized for WWW data 
($\mu = 0.01 \ll 1$) and for AB model at $q=0.1$ ($\mu = 0.1 \ll 1$), while for $q=0.7$
the PageRank is clearly delocalized ($\mu = 0.8 $).

\begin{figure}[htbp]
\begin{center} 
\includegraphics[width=.95\linewidth]{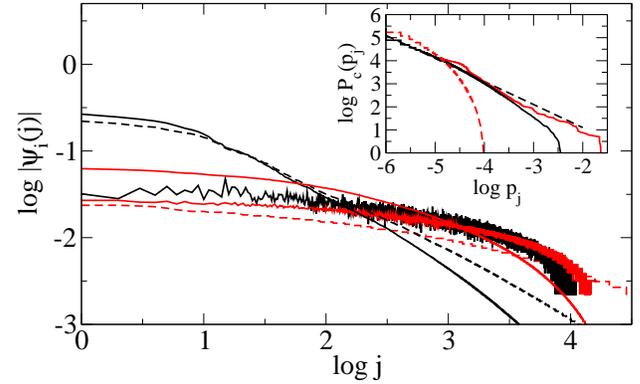}
\end{center} 
\vglue -0.4cm
\caption{(Color online) Dependence of eigenvectors $\psi_i(j)$ of AB model 
on index $j$ ordered in decreasing PageRank values $p_j$ (with normalisation
$\sum_j |\psi_i(j)|^2=1$ and $\sum_j p_j=1$).  Full smooth curves are PageRank vectors
for $N=2^{14}$, dashed smooth curves for $N=2^{19}$.  Non-smooth curves are 
eigenvectors ($N=2^{14}$) within $3<\gamma<4$ with $|\Psi_i(j)|^2$ 
averaged in this interval.
States are averaged over $N_r=5$ random networks. 
Black is for $q=0.1$, red/grey for $q=0.7$.  
Inset: cumulative distribution $P_c(p_j)$ normalized by $P_c(0)=N$
for AB model ($N=2^{18}$ and $N_r=5$)
at $q=0.1$ (full black) and $q=0.7$ (dashed red/grey), and for LJMU non-randomized data
(full red/grey). Dashed straight line indicates slope $1-\nu=-1$. Logarithms are decimal.
}
\label{fig5}
\end{figure}

The distribution of the eigenvector components is shown in Fig.~\ref{fig5} for AB model.
For $q=0.1$ the PageRank is only slightly modified when $N$ is increased by a factor
of $32$ showing a decay $\psi_1(j) \sim j^{-\beta}$ with fitted value $\beta= 0.8$, close
to the WWW value $\beta= 0.9$ \cite{donato}.  The cumulative PageRank distribution $P_c(p_j)$
displayed in the inset also shows a good agreement with WWW data. By contrast, for $q=0.7$, the PageRank
shows a flat distribution over a number of nodes which increases with system size, corresponding
to a delocalization regime.  The states in the bulk are delocalized for both values of $q$.

The obtained results show that  localization properties of the PageRank vector
depend on the type of networks.  Even rather similar networks described by the same
AB model with just one parameter changed show two qualitatively different behaviors.
In one case, which is closer to scale-free networks, 
the localized PageRank is distributed essentially
on a finite number of nodes (finite IPR) while in the other case, closer to small-world type, 
the delocalized PageRank is spread over a number of nodes which grows indefinitely with system size. 
The transition between the two regimes can be viewed as a delocalization transition in the 
Google matrix.  Our studies show that actual WWW networks are located in the localized phase.
The transition to the delocalized phase can drastically affect the efficiency of the Google search.
Indeed, in the delocalized phase the PRA 
still efficiently converges to a well-defined PageRank vector,
which is however homogeneously spread practically over the whole network.  In such a situation
the classification of nodes by PageRank values remains possible but gives almost no significant
information.  We note that this delocalization transition can take place even in presence
of a large gap in the spectrum of the Google matrix.  
The above transition takes place for the PageRank when changing parameters of
the network.  For fixed parameters, we also observe a delocalization transition in the
complex plane of eigenvalues $\lambda$.  
This means that the modes which describe relaxation to the PageRank are generally delocalized
over the whole network for a broad range of relaxation rates $\gamma$.
This transition is reminiscent of the
Anderson transition near the mobility edge in energy eigenvalues. Further studies are required
in order to fully understand the physical origins of these transitions and their dependence
on the characteristics of the networks. 

\vglue -0.30cm

%%**********************************************************************

\end{document}